**All the single cells: single-cell transcriptomics/epigenomics experimental design and analysis considerations for glial biologists**

Katherine E. Prater (ORCID: 0000-0001-8615-207X)[1] & Kevin Z. Lin (ORCID: 0000-0002-1236-9847)[2]

1. Department of Neurology, University of Washington School of Medicine, Seattle 98195
2. Department of Biostatistics, University of Washington, Seattle 98195

Acknowledgements: KEP would like to sincerely thank Drs. Ali Shojaie (UW Biostatistics) and Wei Sun (Fred Hutch Department of Biostatistics) for their helpful input and support on bioinformatics and single-cell analysis pipeline development. The authors would also like to thank Drs. Suman Jayadev, Jonathan R. Weinstein, Jessica E. Young, and Corbin S. C. Johnson for their thoughtful comments and feedback on the manuscript draft.

Word Count: 11,760

Correspondence:
Katherine E. Prater
keprater@uw.edu
1959 NE Pacific St.
Box 356465
Health Sciences Building RR650
Seattle, WA 98195

**Main Points**
- Glial biologists increasingly utilize single-cell 'omics technologies
- The design and analysis of single-cell 'omics is critical to generating reliable insights
- This primer outlines key decision-points and considerations for glial biologists

**Abstract**

Single-cell transcriptomics, epigenomics, and other 'omics applied at single-cell resolution can significantly advance hypotheses and understanding of glial biology. Omics technologies are revealing a large and growing number of new glial cell subtypes, defined by their gene expression profile. These subtypes have significant implications for understanding glial cell function, cell-cell communications, and glia-specific changes between homeostasis and conditions such as neurological disease. For many, the training in how to analyze, interpret, and understand these large datasets has been through reading and understanding literature from other fields like biostatistics. Here, we provide a primer for glial biologists on experimental design and analysis of single-cell RNA-seq datasets. Our goal is to further the understanding of why decisions might be made about datasets and to enhance biologists' ability to interpret and critique their work and the work of others. We review the steps involved in single-cell analysis with a focus on decision points and particular notes for glia. The goal of this primer is to ensure that single-cell 'omics experiments continue to advance glial biology in a rigorous and replicable way.

**Introduction**

Single-cell 'omics studies of various sorts have provided datasets unprecedented in their size and resolution to generate testable hypotheses about glia. One of the significant benefits of these single-cell 'omics approaches is the ability to distinguish and differentiate subpopulations of glial cells. Glial cell types are exquisitely sensitive to their surrounding environment. This results in many distinct phenotypes which each have varied impacts on the surrounding cells. When working with glia, it is then important to describe and test hypotheses about the diversity within these cell types, not just the overall population of all cell types within a donor. For many glial biologists, single-cell/nucleus RNA-seq, multiome (the combination of two 'omics technologies), and various forms of proteomics, lipidomics, and spatial transcriptomics are becoming increasingly relevant to their fields of study. 'Omics approaches provide powerful tools for understanding the biology of glia, and the technologies to generate the datasets in the wet lab are simple enough to use these days. Analysis of these large datasets is also becoming more accessible with a variety of algorithms and toolsets available either open source or through licensed software.

'Omics approaches are descriptive in nature (Diaz-Ortiz & Chen-Plotkin, 2020). Even if one designs a well-controlled multi-group multi-biological replicate experiment, the dataset generated will provide evidence that gene expression or epigenetic markers (or both) changed but will not provide the mechanistic or direct experimental evidence that these changes are biologically meaningful. Validation of the findings by directly testing the alterations in gene expression and what they mean functionally are then required to fully interpret the results of these 'omics datasets.

We write this primer with the goal of providing informed guidelines for researchers on the application and analysis of single-cell 'omics datasets. These sometimes-daunting datasets consist of large matrices of genes by samples by cells. There are new packages available for analysis every day and there is potentially programming to be learned on top of determining what algorithm to apply. These hurdles are all surmountable for glial biologists. We focus on single-cell RNA-seq and single-nucleus multiome approaches with the idea that these are some of the most ubiquitous of datasets to analyze. While our focus is on these specific datasets, the general themes we cover will be broadly applicable to other forms of 'omics technologies and analysis. We provide an overview of the best practices and a guide on how to make decisions when analyzing datasets. Our goal is to make single-cell 'omics more accessible to those interested in applying these powerful tools to their studies. Advancing the science of glial biology will involve knowledge of how/when/why to apply these techniques and the appropriate analysis tools.

## Glossary of Terms

| Term | Definition |
|---|---|
| Differentially expressed gene (DEG) | A DEG is detected when the expression of that gene is significantly altered. DEG analysis typically occurs twice in analysis of single-cell datasets. Once to detect genes that separate clusters (cell types or states), and once to detect genes that differ between treatments or groups. The detection of DEGs for each of these two questions is different as the identification of DEGs that are for groups should involve either a pseudobulk approach or modeling of the source of each cell in the dataset to account for the group-level data. |
| Epigenomics | The class of 'omics technologies used to identify alterations in epigenetics such as open chromatin regions, chromatin conformation, histone modifications, and other forms of epigentic modifications. These modifications can affect gene expression and are both heritable and modified by the environment of an organism. |
| Gene Set Analysis (GSA) | A technique that applies a Fisher's exact test (typically) to a set of significant genes identified by DEG analysis to detect biological pathways or other correlated biological functions with those genes. This analysis takes only the significant genes, does not account for their directionality or the magnitude of effect, and is not well powered when gene lists are small (<100). |
| Gene Set Enrichment Analysis (GSEA) | A technique that identifies correlated biological pathways using the entire list of highly variable genes (or the full genome) and combines them into enriched pathways using both the direction and magnitude of change in the dataset. This pathway analysis technique is preferred because it takes into account the fold change of the genes and detects where groups of genes are changing in a correlated manner. Note that the genes identified as contributing to the pathway enrichment may not be significantly different themselves. |
| Highly-variable gene | A gene whose expression level varies significantly across different cells in the dataset. In some cells it has very high expression and in other cells it is lowly expressed. Computationally identifying highly-variable genes helps to focus the analysis on the most informative parts of the data. |
| Library | A unit of cDNA prepared for sequencing through appropriate size discrimination and sequencing platform adaptor ligation. A single library typically contains all the cDNA transcripts from a single source. Libraries are typically prepared in batches of samples, which are a source of batch effects in the sequencing dataset. |
| Model | A statistical equation to describe the datatset. An example of "modeling" is to determine differential gene expression while accounting for donor-level covariates. |
| Multiome/multi-omics | The term used when multiple forms of 'omics technologies are applied to a single cell source. This is often used to refer to the combination of RNA-seq and ATAC-seq performed on the same nucleus, but can refer to any combination of two or more 'omics technologies performed on the same cells/nuclei. |
| Multiplexing | A technique used to label multiple sample sources such that they can be effectively pooled for single cell separation and library preparation. The labels allow the sequence from each sample source to be later identified as a separate sample in the dataset. This technique reduces reagent costs but allows use of the full N of samples obtained, a better outcome than pooling without multiplexing. |
| Over-dispersion | Variability in the sequencing count that is greater than expected based on a given statistical model. |
| Principal component analysis (PCA) | A technique that reduces the number of "dimensions"/"features" in your data while keeping the most important information. It helps you see patterns and trends by creating new variables (called principal components) that are combinations of the original variables, but fewer in number. This makes complex data easier to visualize and analyze. PCA falls in a category of methods called "dimension reduction methods", of which PCA is one of the most common methods for this task. |
| Proteomics | The class of 'omics technologies used to identify changes in protein expression within a given cell source. |
| Pseudobulk | A method in single-cell RNA-seq analysis where the data from individual cells generated from the same cell source are aggregated or summed to create a bulk RNA-seq-like profile for that source. This approach helps to mitigate the noise and sparsity inherent in single-cell data by combining the expression levels of cells, making it easier to apply traditional bulk RNA-seq analysis methods. |
| Sequencing count | The non-negative number (e.g., 0, 1, 2, ...) representing the number of fragments from a particular gene sequenced for a particular cell. The single-cell sequencing matrix is typically called a "count matrix", and is oriented to have one row per gene, and one column per cell. This matrix is often sparse, meaning many of its entries are 0 (no counts for a given gene for a given cell). This sparsity needs to be addressed by the statistical applications developed for single-cell analysis. |
| Sparse | A matrix is sparse when it contains cells that are zero. In single-cell datasets, there are many zeros because not every cell contains counts sequenced for every gene in the genome. |
| t-SNE/UMAP | Two dimension reduction methods that are particularly well-suited for the visualization of high-dimensional datasets. The t-stochastic neighbor embedding (t-SNE) and uniform manifold approximation and projection (UMAP) are the most common ways to visualize single-cell datasets. While these two methods offer more insightful visualizations of single-cell sequencing data within one plot when compared to PCA, the coordinate-system produced in these plots should not be over-interpreted. |
| Type-1 error | Also known as a "false positive," this occurs when a statistical test incorrectly rejects a true null hypothesis. In other words, it is the mistake of concluding that there is an effect or difference when, in fact, none exists. A "Type-1 error inflation" means that there are substantially more incorrect rejections than anticipated. |
| HI-C | An epigenomics proximity ligation method that captures the organizational structure of chromatin in three dimensions, where genomic sequences that are distal to each other in linear terms are close to each other in 3D space. |

## Designing the experiment:

**Why use single-cell/nucleus transcriptomics?**

If the scientific question being asked is about how cells might shift their gene expression in response to a specific condition/disease state/stimulus and specifically involves understanding the diversity of responses within the larger cell population, then single-cell transcriptomics may be the right approach. If the question is whether gene expression changes overall within a well-defined cell type that can be enriched for and isn't suspected to have a high degree of heterogeneity or a whole organ, then bulk RNA-seq is the technique of choice. The power of single-cell datasets is primarily in detecting unique shifts in different subpopulations of cells rather than a shift in gene expression at the larger population level. Single-cell RNA-sequencing technologies can be applied to either whole cells or nuclei. For the remainder of this primer, we will refer to cells when the word is generic, though know that this could be cells or nuclei depending on your application. This allows their application to a wide range of tissues including fresh, fresh frozen, and fixed from which to extract glia and analyze their gene expression. Bulk RNA-seq is typically much cheaper, potentially faster than single-cell approaches, more sensitive to detect lowly expressed genes, and enables better capture of splice isoforms. However, technologies continue to improve that may allow for the low-cost processing of hundreds of single-cell samples for transcriptomics analysis in the future and technologies to merge long-read sequencing with single cell approaches will improve the power of single-cell sequencing to identify splicing and more rare transcripts.

**Why use single-nucleus epigenomics?**

If the scientific question involves the chromatin accessibility, conformation, or histone modifications that shift gene expression, then single-cell epigenomics is a likely technique candidate. The assay for transposase-accessible chromatin with sequencing (ATAC-seq) detects regions of open chromatin. Single-nucleus methylation technologies are also available if epigenetic markers driven by methylation are of interest for the science. Single-cell Hi-C allows an understanding of the chromosome conformation (Flyamer et al., 2017; Nagano et al., 2013). Single-cell cut and tag identifies histone modifications and areas of open chromatin (Bartosovic et al., 2021; Bartosovic & Castelo-Branco, 2023). In the case of ATAC-seq datasets, it is often useful to have a reference dataset of single-nucleus RNA-seq to assist with clustering the dataset to then determine the cluster-level peaks of open chromatin regions. Note that ATAC-seq requires nuclei, so while transcriptomics at the RNA level can use cells or nuclei, the technology behind epigenomics requires a shift to nuclei.

**Why use single-nucleus multiomics?**

Multiome is the combination of 'omics technologies from the same cell source. Single-nucleus multiome is commonly used to refer to the combination of performing RNA-seq and ATAC-seq on material generated from the same nucleus. There are additional single-cell multi-omic technologies that allow combinations of lipidomics, proteomics, HI-C, or DNA methylomics, but these technologies are still emerging. Colloquial use of "multiome" may shift as technologies beyond the combination of RNA-seq and ATAC-seq become more widely used. The power of performing multiomics is that both types of 'omics datasets are generated simultaneously. In the case of single-nuclei RNA-seq and ATAC-seq, this would allow one to correlate a specific gene expression change to differential accessibility of its regulatory region within the same cell. Multiome avoids the need to integrate datasets from different physical samples, and the issues of needing to identify a good reference dataset as both datasets are built from the same nuclei. Several publications are out or forthcoming with glial multiome datasets (Trevino et al., 2021; H.-L. V. Wang et al., 2023; Xiong et al., 2023; Zhao et al., 2024). We also note there are promising methods that help "predict" the ATAC-seq data from RNA-seq data or vice-versa (Cao & Gao, 2022; Gong et al., 2021), as well as methods to computationally link RNA-seq profiles from one sequencing run to ATAC-seq profiles from a different sequencing run (Chen et al., 2023), but it is still not clear if these methods work for all biological systems related to glial cells.

**Questions regarding splice variants or small non-coding RNAs**

In cases where investigating splice variants or small non-coding RNAs like microRNAs is the goal, specialized technologies need to be utilized. The standard single-cell RNA-seq protocols will not be sufficient. In short-read sequencing (the traditional single-cell RNA-seq technology), the cDNA is cut to a specific fragment length near either the 3' or 5' end of the transcript. Thus, sequencing does not provide a whole transcript, but simply enough bases near one end of the transcript to identify the gene associated with the transcript in question with some specificity. These short reads typically do not provide datasets sufficient to answer scientific questions regarding small RNAs or splice variants outside of the small region of transcript.

To identify whether specific splice variants are present in subpopulations of glial cells, then single-cell long-read sequencing approaches should be considered. These approaches allow for sequencing of the full-length (or close to it) transcript, whereas traditional 3' or 5' single-cell RNA-seq will not capture variations in the transcript unless they are very close to the 3' or 5' end. Nonetheless, researchers have discovered some glial splice variants on the 3' end using short-read single-cell RNA-seq sequencing (Fansler et al., 2024; Gao et al., 2021; Kang et al., 2023), but this likely represents only a fraction of the discoveries that will be made in the upcoming years with long-read sequencing. Single-cell long-read sequencing has advanced significantly in the last few years, so the ability to detect splice variants and other variations in transcript makeup are more available now.

There are also now several technologies available to sequence small RNAs like micro-RNAs from single-cells (Benesova et al., 2021). While not as widely utilized as standard short-read single-cell RNA-seq, they exist and are available for use. Several studies have applied small RNA-sequencing to glia to identify unique results in micro-RNA regulation of glial function (Huang et al., 2023; Li et al., 2022). The application of these alternative technologies, while imperative for asking these scientific questions, may require alterations in the analysis approaches described below. In this primer we focus on traditional short-read single-cell RNA-sequencing approaches as they are the most common.

**Enriching datasets for glial populations of interest**

One consideration for the experimental design is whether to enrich for cell types of interest prior to performing single-cell 'omics. This may be particularly important when working with glia, as their populations are often more limited compared to other cell types. For example, microglia are estimated to make up 5-10% of human brain cells (Blinkow & Glezer, 1968; Pelvig et al., 2008). However, in single-nucleus datasets isolated directly from human brain they make up just 2-3% of the nuclei that can be analyzed (Alsema et al., 2020; Del-Aguila et al., 2019; Mathys et al., 2019; Olah et al., 2020). This sparsity in the number of microglia nuclei available for analysis limits the ability to capture the complete representation of the full diversity of subpopulations of this cell. Datasets in both human and mouse with limited numbers of microglia detect fewer subpopulations than datasets that enrich for microglia (Gerrits et al., 2021; Mathys et al., 2017, 2019; Nguyen et al., 2020; Prater et al., 2023). Fortunately, techniques exist to enrich for all types of glia ahead of performing single-cell 'omics assays that are easily employed and will make the resulting dataset more powerful for answering specific scientific questions (e.g. Gerrits et al., 2021; Nott et al., 2019; Ochocka et al., 2021; Prater et al., 2023; Sadick et al., 2022; Schroeter et al., 2021; Wei et al., 2023; Yang et al., 2022). Enrichment can be achieved through positive selection (by using a protein marker on your cell type of interest), or via negative selection (by using a protein marker on alternative cell(s) to remove them from the dataset). Both are viable strategies to enrich for specific glia in the dataset and each has a different set of caveats associated with it. For example, positive selection may miss selecting for subpopulations where the protein used to select was either not expressed or expressed at low or undetectable levels. Negative selection may not enrich for a particular cell type as effectively. Examples of both positive and negative selection can be found in the literature for many types of glia.

**Determining sample size**

An important aspect of experimental design is how many samples to collect. For the purposes of this discussion, we will refer to technical replicates as replicates of a single sample that may be generated from either multiple cultures, multiple differentiations, or multiple cell isolations from a single source (cell line or animal). Biological replicates, in contrast, are replicates from multiple different sources, whether those are unique individuals (animals or human), or cell lines. There are ways of calculating power for single-cell analyses (see Jeon et al., 2023 for a review), which should assist in the design of studies. In general, the answer (as always with statistics) is that more samples will give you more power. If your experimental question is comparing groups, then it will be necessary to have more than one biological replicate per group. There are scientific questions where one might ask about shifts in cells within a sample itself – this may require multiple technical replicates but potentially fewer biological replicates, although for validity it would likely be better to have multiple of both. While there is no specific number of samples that can be given, the need for multiple biological and technical replicates depending on the scientific question being asked is critical, and datasets with a single data point should be viewed with caution since their generalizability is unknown.

One discussion point of note is that no matter whether technical or biological replicates are generated, pooling those replicates into a single sequencing library (the unit of transcripts prepared for sequencing) without multiplexing reduces the N to one. This compresses the replicates together and reduces the dataset to a single point, an undesirable outcome. Better alternatives are multiplexing technologies which allow researchers to pool their samples for cost-efficiency. Multiplexing typically involves attaching an identifier (potentially an oligonucleotide or some other tag) to a specific sample and then pooling samples with different tags for the generation of the single-cell library. Once the library is sequenced, the identifiers can be used to de-multiplex or to pull the samples back apart so that the higher sample number can be used. This is very different from pooling samples and then receiving a single sequencing file back which cannot determine which cells came from which sample. The latter case would be a dataset from a single data point rather than a multiplexed dataset which allows the use of the replicates as multiple individual data points. There are many published studies where replicates were collected and then pooled without multiplexing, which effectively reduces the power of the study. We strongly encourage researchers to use multiplexing if pooling must occur so that the generation of replicates can be utilized fully in the datasets.

**Cells or nuclei, which is better?**

In some cases, like when using ATAC-seq or multiomics technologies, researchers are required to use nuclei. However, in traditional single-cell RNA-seq, researchers have the choice of using the technology either on single cells or single nuclei. In some cases, the choice of whether a study is performed on cells or nuclei is driven by the tissue available. For example, it is currently extremely difficult to isolate whole cells from flash frozen human brain tissue. Thus, if a study is utilizing archival samples, it is likely that they are using single nuclei. In contrast, freshly resected tissue is more easily dissociated and allows for the isolation of whole glial cells for single-cell RNA-seq. There are studies which suggest that the transcriptional profile of single nuclei may differ from that of whole cells (Thrupp et al., 2020). However, there are additional studies which suggest that the transcriptional profile may not differ as widely (Lake et al., 2017). Regardless, there is a clear consensus that the dissociation is critical while collecting single cells from fresh tissue to avoid inducing spurious gene expression profiles that are altered by the processing of the tissue itself (e.g. Marsh et al., 2022; Mattei et al., 2020).

**Batching in an experiment**

One piece of critical experimental design consideration is how to set up library and sequencing batches. In this instance, a batch is a set of samples that will have their libraries or sequencing generated together in one grouping. We will discuss batch correction further in the analysis portion of this primer as it is an important component of analysis as well. Multiple studies have identified batch effects in large studies of sequence (Hartl & Gao, 2020; Katayama et al., 2019; Lauss et al., 2013; Leek et al., 2010; Taub et al., 2010; Tung et al., 2017). These studies demonstrate that counterbalancing and consideration of batches (how samples are grouped) in the experimental design is critical to the results of the study. Studies using single-cell technologies typically involve generating multiple rounds of library batches. While minimizing batches where possible is useful, there is often no way to avoid multiple library batches in particular because of processing time and sample expiration dates. What is critical is ensuring that, where possible, these batches are not confounded by biological variables of interest (Hartl & Gao, 2020; Lauss et al., 2013; Leek et al., 2010; Tung et al., 2017). For example, a library batch should have equal representation of "cases" and "controls" or as close to it as possible. If library or sequencing batches are represented primarily of a single experimental cohort, the batch confound and the biological variable of interest will be unable to be fully disentangled and result in an inability to fully interpret the results of the study. It is almost impossible to avoid batch effects and we'll discuss their removal during the analysis section of this primer. However, counterbalancing and randomizing samples across library and sequencing batches is critical to ensure high quality datasets.

**Study Design Conclusions**

Once you determine that you will use single-cell 'omics of whatever sort best fits the experimental question, then the goal will be to design the experiment to minimize batch effects and utilize the appropriate number of samples. Power analysis can help determine an appropriate number of samples needed to answer the questions of interest. As noted above, we caution against pooling samples without multiplexing as this practice removes the ability to understand biological variability in the dataset. Practicing appropriate counterbalancing of samples in the scheduling of experiments is also critical to avoid introducing batch effects in library preparation or sequencing that will cause the dataset to be uninterpretable. Many of these practices are standard in non-single-cell experimental design as well, so all that is required is a simple transfer of applicable skills to these new experiments.

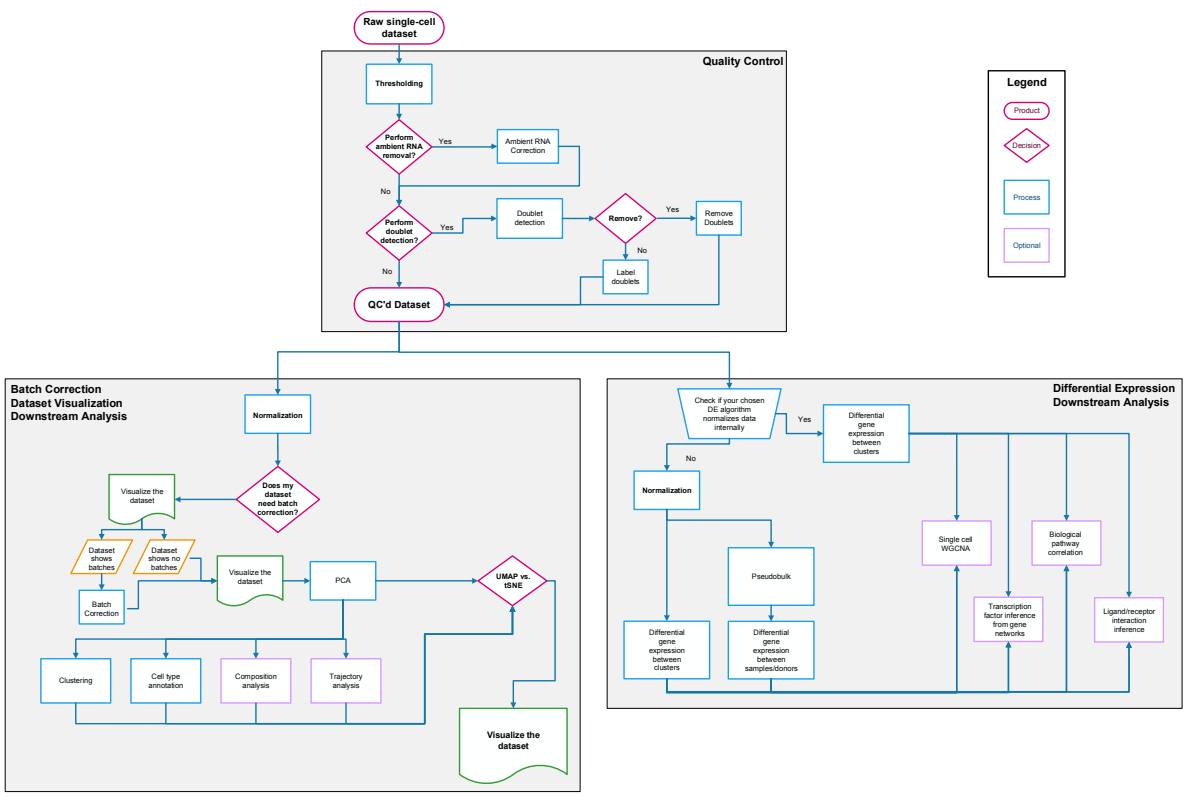

***Figure 1 caption***: Overview of single-cell/nucleus RNA-seq analysis. We provide a flowchart workflow for standard single-cell/nucleus RNA-seq analyses. Analysis begins with a raw dataset that undergoes quality control (QC) which includes thresholding and may optionally include the removal of ambient RNA and doublets. From the QC'd dataset analysis can proceed to batch correction and downstream methods that rely on more on data visualization. The other path is to determine differentially expressed genes, which may include normalizing the dataset but should not be performed on the batch corrected dataset. In either the batch correction or differential expression output areas of analysis there are several optional analyses (in purple boxes) that can be applied to datasets. The analysis options in blue are standard approaches seen in the field.

**Analysis:**

Once a well-designed study is completed next comes the analysis. These large datasets require significant amounts of computing power and usually some ability to script in a language like R or Python. Several groups have generated single-cell transcriptomics analysis tutorials that are excellent resources for specific packages and the analysis steps that can be applied to datasets (Amezquita et al., 2020; Andrews et al., 2021; Haque et al., 2017; Heumos et al., 2023; Lähnemann et al., 2020; M. Wang et al., 2022).

Not every dataset is alike, and the algorithms available for single-cell analysis change rapidly. The tools that worked on a previous dataset may not work as well on a new one. New better tools may be available but will require troubleshooting and assessment to confirm they work well for the current dataset. When choosing packages, we recommend selecting tools that are commonly seen in publications in your field. We also recommend identifying GitHub repositories for packages that have active updates and responses to issues. This suggests that the tool is actively being maintained and you will be able to find support from the authors of the tool.

Below we outline a general approach to analysis for single-cell datasets and identify where the approach would differ for multiome datasets (which includes analysis of single-nucleus ATAC-seq data). Our goal is to outline important decisions and why you might choose one approach over another (see **Figure 1**). There are a multitude of packages and options for approaching single-cell analysis written in both R and Python. Where important, we make specific package recommendations, but our intention here is to outline how to think about the application of analyses to datasets rather than to identify specific packages to use since packages can change regularly. Our goal is to leave readers more informed about how to approach datasets, both their own and others', so that the use of 'omics in glial biology can continue to enhance our understanding and build trustworthy new hypotheses to test.

**One important note about single-cell analysis**: Traditional statistical analysis does not typically involve visualizing a dataset to make informed decisions. In fact, traditionally visualizing a dataset and making decisions based upon what is seen leads to inaccurate and faulty statistical analysis. Single-cell analysis is the exact opposite of this traditional method. There are many times when the best way to determine whether the algorithm applied to the dataset was effective is to visualize the dataset and see how it looks. You'll note throughout this primer that we recommend visualizing your dataset (and we discuss what visualizations are commonly used) and that is why. There may be a future where visualizing the dataset in between statistical applications to the dataset becomes unnecessary because there are metrics and quantitative ways to interpret the results, but currently the field is set up such that visualization is part of the way decisions are made as to how to approach these data.

**Quality control**

One of the first steps in analysis of single-cell datasets is quality control (QC). This includes setting some thresholds to exclude poor quality cells from the analysis, as well as doublet and ambient RNA identification.

Setting thresholds is typically the minimum QC for a single-cell dataset. Thresholds are set to remove cells with low quality sequence and low numbers of genes expressed. Ideally this removes the empty droplets generated by separating the single cells. Setting thresholds for the dataset is a critical step in analysis because it can exclude cells with quality sequencing if done too stringently. For example, microglia often have a smaller number of unique transcripts than other cell types, likely because of their biological size. A higher threshold on the number of unique transcripts per cell can potentially exclude microglia from the dataset being analyzed not because they're low quality, but because they as a cell type express fewer transcripts. Alternatively, settings thresholds too loosely can allow empty droplets or poor quality cells into the analysis which may make later algorithms struggle. Identifying the location for each sample where the majority of empty droplets are excluded but one is unlikely to exclude high quality low expressing cells is key.

Doublets occur in processing single-cell datasets when more than one cell is captured in a droplet or capture area and transcribed with the same barcode as if it were a single cell. In the past, an upper bound threshold on the number of genes expressed was set to exclude doublets from analysis by excluding cells that had very high transcripts. Newer technologies provide ways to identify doublets algorithmically based on their gene expression rather than on an upper bound threshold (e.g. McGinnis et al., 2019; Wolock et al., 2019). These technologies are far superior to the thresholding techniques originally applied, and we highly recommend the use of a package to identify doublets. Recently, researchers are moving toward applying two or three doublet-calling packages and then removing doublets that were identified in a union of the package results. This approach gives confidence since multiple slightly different algorithms have identified a cell as a doublet. One consideration for quality control of datasets is whether to fully remove doublets from the dataset early on, or simply to mark them as doublets with an identifier to use later in analysis. One advantage of the labeling approach is that you continue to visualize the cells labeled as doublets in your later analysis techniques and can determine whether other cells may have been missed in the labeling process. However, leaving doublets in the analysis can also potentially confuse results later on, so removing high confidence doublets remains a reasonable and favored approach to analyzing a dataset.

   Ambient RNA is another confounder and place where quality control can occur in the analysis early on. Ambient RNA contamination occurs in single-cell technologies because RNAse-inhibitors are present and RNA that exists in the solution from lysed cells is sequenced as if it were transcript from the cell of interest. Tools exist to identify transcripts likely to be ambient in the dataset and are particularly useful in single-nuclei datasets where ambient RNAs are likely to be common (Fleming et al., 2023; Young & Behjati, 2020). The presence of ambient RNA can alter downstream analyses like differential gene expression where genes that may have differential expression are not identified because they are weakly expressed across the whole dataset. However, some researchers argue that the algorithms developed for ambient

RNA removal are too stringent and remove more than they should from datasets. These criticisms are valid and so this approach should be carefully applied and reviewed for each dataset. In general, ambient RNA removal is recommended for single-cell datasets, but not all researchers will agree with this. We recommend running your analysis with and without ambient RNA removal. Ideally, the removal of ambient RNA only increases the power of your analyses but does not yield drastic and conflicting results compared to the analysis without ambient RNA removal. If the ambient RNA removal results in drastically different results, we recommend using the analysis without ambient RNA removal. In the more likely scenario that the two are similar but the p-value thresholds are slightly higher in the corrected (ambient RNA removed) dataset, we recommend proceeding with the corrected dataset.

**Normalization**

After quality control, the next step is to transform the data into a more biologically meaningful proxy of gene expression. This step is called normalization because it provides the relative ratio of expression, not the absolute expression. Normalization is similar conceptually to normalization in other methods (like western blotting or qPCR where data is normalized to housekeeping proteins/genes) but is specific for the sparse nature of single-cell RNA-seq datasets. Remember that the count matrix associated with a single-cell RNA-seq dataset is sparse or contains many zeros (no counts for a particular gene for a particular cell). Normalization methods therefore need to account for the "count nature" of sequencing data. The "counts", or number of times a gene has been sequenced during a run, can be present or absent in a given cell due to both biological and technical factors. Factors beyond biological processes contribute to the variability in gene expression in these datasets. Technical contributions to variability include differences in the amount or efficiency of sequencing in cells or contribution of genes that are not biologically informative. To isolate the biological drivers of gene expression changes, we first try to remove the technical effects. Because there can be variability between cells, even from the same sample, of the number of times a gene is sequenced due to technical reasons, there is a need to normalize the "depth" or number of times a gene is sequenced, to the overall number of times all genes are sequenced.

The preferred normalization procedure has evolved dramatically over the years due to the simultaneous advancements in sequencing technology and statistical modeling. Two key ideas have stood the test of time and are found in most modern normalization procedures: the first is to model the ratio between a cell's count for a specific gene and the total counts for the cell (i.e., the sequencing depth for that cell), and the second is to model the over-dispersed nature of single-cell sequencing counts (see glossary). In the earlier days of single-cell RNA-seq analysis, these two aspects were solved by computing the sequencing depth of a cell and then performing a log transformation. This procedure is called log-normalization. However, more recent papers have demonstrated that this simple transformation often over-represents cells with a small but non-zero count (Hafemeister & Satija, 2019; Townes et al., 2019). Hence, methods that explicitly model the count nature of the data, are now more widely used (Hafemeister & Satija, 2019; Lopez et al., 2018). We note that benchmarking papers provide slightly different recommendations for normalization (Ahlmann-Eltze & Huber, 2023; Choudhary & Satija, 2022; Lause et al., 2021). Overall, we recommend you choose a diagnostic metric (a quantitative value or visual criterion) to check prior to normalizing the data, and then choose the normalization method that best maximizes that metric. This is because while the benchmarking papers offer a good overview of all the normalization methods, the authors' recommendations might not apply to your particular biological system, sequencing technology, or available budget for sequencing depth. We have found the diagnostic metrics that check how gene expression correlates with each cell's sequencing depth or how much variation is explained by genes of different mean sequencing depth to be quite useful (Hafemeister & Satija, 2019).

Newer normalization methods have the added benefit that confounding covariates can be adjusted for (i.e., "regressed" out). This ensures that the resulting normalized dataset for downstream analyses is free of biological effects that might obfuscate the intended biology you're studying. Two covariates commonly adjusted for are the cell's percentage of gene expression from mitochondrial genes, which is increased in stressed or dying cells, or the cell's cell-cycling phase. However, if you're working with human or animal cohorts, there is an additional concern of how you would remove the donor-level covariates such as age, sex, cognitive score, etc. Since each donor contributes many cells, all the cells from the same donor have the same donor-level covariate. This means that methods like SCTransform that normalize each gene separately struggle to adjust for donor covariates. (For example, it isn't easy to reliably estimate the effect of age when there are not many different values of donor ages and also a lot of zeros in a gene's expression.) Alternatively, for scenarios like this, normalization procedures that group information across all the genes (e.g. scVI or GLM-PCA) when accounting for donor covariates are recommended (Lopez et al., 2018; Townes et al., 2019). These methods come with an added benefit of providing a low-dimensional embedding directly, which will help with later batch correction, cell type labeling, and visualizations. Regardless, we recommend visualizing the data with and without removing donor covariates to gauge how successful a normalization procedure was.

Since normalization methods can struggle with donor-level covariates, it is commonplace to perform a principal component analysis (PCA) afterwards since PCA will additionally help to reduce technical variability. At a high level, a PCA defines a low-dimensional space driven mainly by the densely sequenced genes. A PCA provides a short vector to describe each cell by the number of PCs, instead of a long vector of the counts of every gene expressed by that cell over the whole genome. This condensing of signal is how a PCA can help further clean up your single-cell dataset. Many methods then use the PCA result for visualization or further analyses.

After the normalization, it is also commonplace to assess which genes are "highly variable." All downstream analysis is then performed only on these highly variable genes. Typically, most analyses use somewhere between 2,000 to 5,000 genes. The choice of 2,000 or 5,000 is often based on computational resources and time available to analyze a dataset. These highly variable genes are selected computationally because their variability across all the cells is large relative to the mean normalized expression. Many computational tools we will discuss later treat each gene equally. Too many uninformative or noisy genes in the analysis might hinder the performance of these downstream computational tools, which is why the dataset is limited when applying these.

One last note – of all the different analyses for a sequencing dataset, normalization is typically the step that is 'omics-technology-specific. The normalization procedures for single-cell RNA-seq are different from those for single-cell ATAC-seq, long-read sequencing, etc. As we discuss in the next section, this is complicated since some batch correction and differential analysis methods involve normalizing the data within the method implicitly. Single-cell tutorials (e.g. Amezquita et al., 2020; Andrews et al., 2021; Haque et al., 2017; Heumos et al., 2023; Lähnemann et al., 2020; M. Wang et al., 2022) are a good place to start to determine what method will be best for your particular application.

**Batch Correction**

Earlier we mentioned the need for batch correction in many single-cell datasets. While not necessary in all cases, batch correction is often beneficial (or needed) to account for artifacts introduced by technical confounds such as library preparations or sequencing batches (Leek et al., 2010; Tung et al., 2017). The need for batch correction can often be identified by visualizing the dataset (more on this later) and identifying whether the dataset appears striped or there is a clear separation of the dataset along axes of potential confounds such as sample, or sequencing batch, or library preparation, when looking at the colorized display of cells. A very large batch effect will result in the data separating into two portions of the visualization based on a technical confounder. More often, the batch effects are more subtle and may simply result in mild striping or small differences in the visualization of the dataset. If your dataset from all conditions, sequencing batches, etc. perfectly overlays or is nicely mixed in the great majority of areas of your visualization then batch correction may not be necessary.

Once you have determined that batch correction is needed, there are multiple tools that could be implemented to complete this step. We should note that it's possible that different batch correction methods may perform differently on different tissue types (Luecken et al., 2022). Batch correction methods can take several forms which have been overviewed in many benchmarking papers (Chu et al., 2022; Luecken et al., 2022; Tran et al., 2020). While over-correcting during batch correction can mean the loss of a true positive in cell state presence or absence, under-correcting could lead to false results of a difference between groups if groups are confounded by batch effects. As discussed earlier, it is incredibly important that samples of different groupings be counterbalanced as evenly as possible across batches. When batch is confounded by a biological variable of interest, the results cannot be interpreted properly (Hartl & Gao, 2020). **Figure 2** depicts one batch correction using scVI (Lopez et al., 2018) applied to two hypothetical datasets where one has sequencing batches perfectly aligned with the biological variable of interest, while the other dataset has the biological variable of interest balanced between the two batches. The figure demonstrates that when there is perfect alignment between sequencing batches and the biological variable of interest, it is impossible to assess if the batch correction was properly done or if there are significant biological differences. This showcases that you should carefully determine how to balance your samples when sequencing your cells.

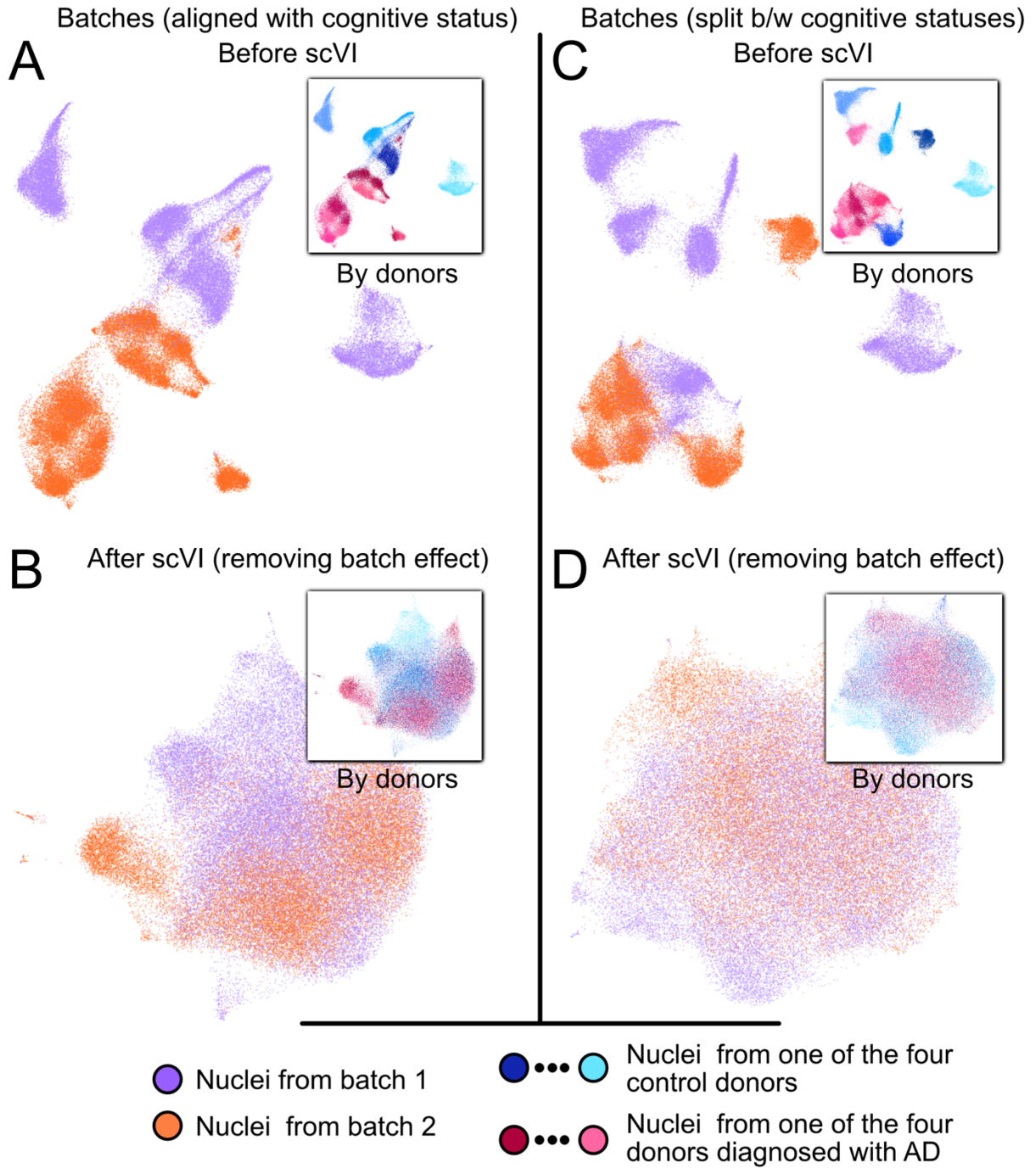

*Figure 2 caption*: Impact of batch correction and difficulties when batch is aligned with the phenotype/treatment of interest. A) In a hypothetical dataset where the batches (in purple or orange) are perfectly aligned with donor phenotype (shown in the inset, where each shade denotes the donor of a nucleus and blue or red denote the two different phenotypes). In this example, phenotype is the cognitive status of the individual and batches contained one or the

*other phenotype but not both. It is normal to see separation of data by donor and/or batch prior to batch correction. B) After applying batch correction, we see a modest mixing of nuclei by batch, but not much mixing by donor phenotype. We cannot disentangle if this modest performance is due to significant biological differences between the two classes of donors or because we did not apply a good batch correction method. C) Another hypothetical dataset where the donor phenotypes are equally split between the batches. D) After applying batch correction, we see a substantial mixing among nuclei across the two batches but there are still substantial differences between the two phenotypes of donors. All the plots shown are UMAPs where each point is a nucleus, and originated from a dataset of microglia among donors with and without AD (Prater et al., 2023). scVI is used here as an example of batch correction.*

Batch correction is one of the more important steps that can be applied to datasets for appropriate visualization and downstream analyses based on corrected data. We recommend applying a standard batch correction tool often used in your field and visualizing your results. If you so choose, you could apply several batch correction methods to determine whether they may be correcting the dataset in different ways. Tran et al. (2020) and Korsunsky et al. (2019) offer suggestions on how to assess which batch correction method worked the best. Typically, this involves computationally quantifying if there is a good mixing of cells across the different batches.

**Data visualization**

Most single-cell papers have a so-called "UMAP plot" or "t-SNE plot" to visualize the single-cell data as demonstrated previously in **Figure 2**. The intent of such plots is usually to provide the reader with a quick bird's-eye summary of how many cells the dataset contains, how many cell types, and how separated the cell types are. In these plots, each point is a cell/nucleus, and the color of each point identifies the cluster. Clusters most commonly represent the cell type, cell cluster, or cell sub-state (more on this in the following section). If you have donor information or batches, you might also color your cells based on that covariate.

Alternatively, you might color each point according to a color gradient based on how high or low that cell's gene expression is. These visualizations are highly flexible and quite powerful. As mentioned earlier, visualizing the dataset is particularly critical in single-cell analyses as the way the visualization looks is often a decision point in assessing whether a given algorithm in the analysis was effective.

The computation of t-SNE (Maaten & Hinton, 2008) or UMAP (Becht et al., 2019) is typically based on the PCA embedding. Technically, both PCA embedding and t-SNE/UMAP are forms of "dimension reduction." However, the key difference lies in the number of dimensions they retain. A PCA embedding usually retains many dimensions, typically around 30 and are used for many methods in a downstream analysis. In contrast, t-SNE and UMAP are specifically designed for visualization, and therefore, they retain only two dimensions. This distinction is crucial to understanding the role of dimension reduction in visualization.

It is important to know that t-SNE and UMAP have some randomness inherent in them since they visualize high-dimensional data. If you have 5,000 highly variable genes, your visualization condenses all those genes into a two-dimensional plot. A lot of distortions are expected. A good analogy to think about is maps of Earth. If the Mercator projection is used to visualize the three-dimensional world in two dimensions, area is distorted (Greenland deceivingly looks as large as Africa) and distance is distorted (the distance from California to Japan deceivingly looks twice as much as the distance from France to Japan). These distortions are even more exacerbated when 5,000 genes are visualized in two dimensions. Nonetheless, UMAPs and t-SNEs are still vital tools for assessing the dataset. In general, UMAPs are often preferred over t-SNEs since UMAPs (empirically) better capture the structure of the dataset. In general, both these visualizations can capture "large cell type separations," but the literal distance between cell types based on these plots should not be over-interpreted. You can use tools to numerically check how far apart cells really are (Johnson et al., 2022; Xia et al., 2024). **Figure 3** depicts different visualizations of glial cells from the same single-nuclei RNA-seq dataset where the cell states are labeled. We can appreciate that the coordinates of the UMAP and t-SNE are arbitrary, as different instances of computing UMAP or t-SNE yields different visualizations. Despite this weakness, these plots contribute useful information by summarizing the differences between cell types and cell states all in one plot. In contrast, PCA coordinates are well-defined, meaning every time you compute the PCA, you will get the same result. However, the linearity of PCA means it cannot separate the cell types or cell states when you visualize two principal components at a time unless you make many plots.

As a rule of thumb, while UMAPs and t-SNEs sometimes have a contentious reputation for providing misleading insight (Chari & Pachter, 2023), these tools still play an essential role in single-cell analyses because a "perfect" visualization of high-dimensional data will never exist. While quantitative tools such as clustering and differentially expressed gene (DEG) analyses offer insight by themselves, applying qualitative tools such as visualizations is equally essential. If your quantitative analysis uncovers a strong pattern among the cells, then, you should see some qualitative evidence of the same signal when visualizing the data appropriately, and vice-versa.

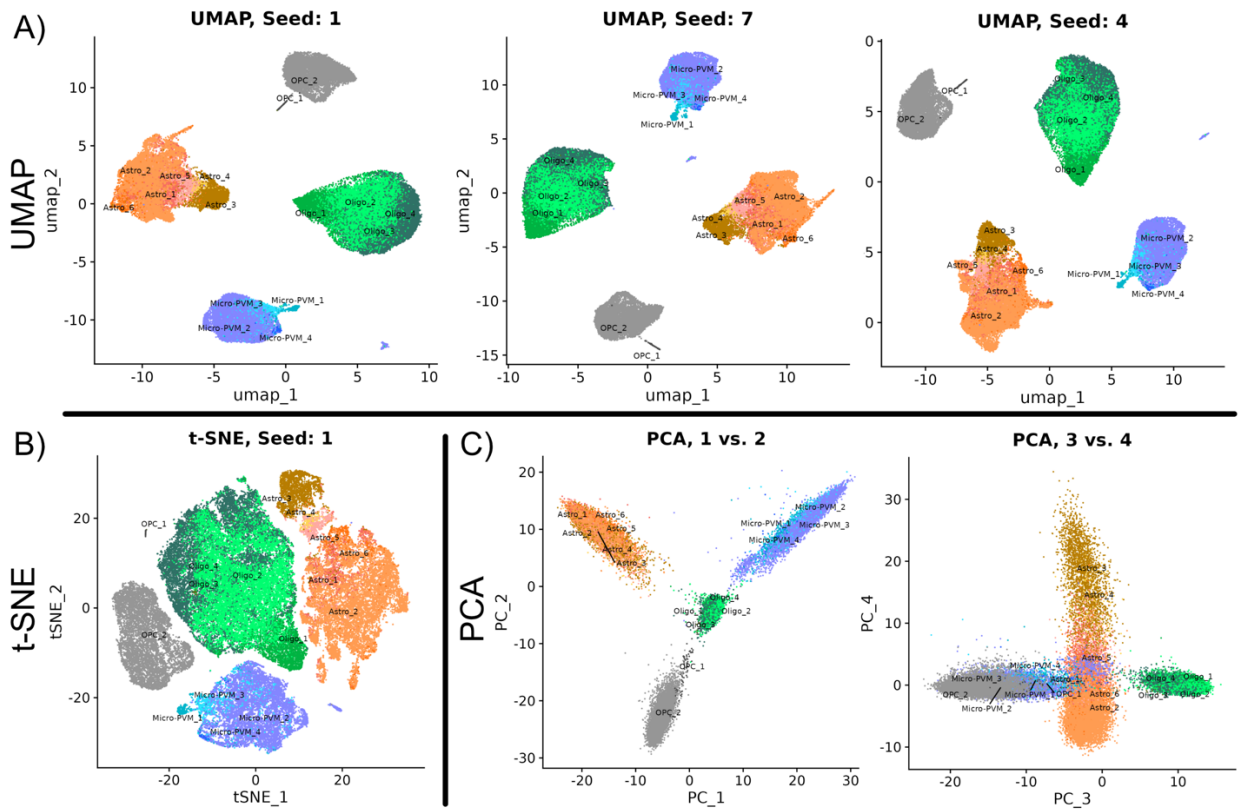

*Figure 3 caption*: *Differences in visualization of a single-nucleus RNA-seq dataset. A) Three different UMAPs which were computed using the same dataset and parameters. This demonstrates that every time you compute a UMAP, you might get a slightly different plot. Furthermore, the orientation of the cell types and distances between cell types is meaningless in a UMAP, so the UMAP mainly offers a glimpse into the number of cell types and a rough dichotomy of cell states within each cell type. B) A t-SNE plot, which usually shows less separation among cell types compared to a UMAP but typically offers greater spread to appreciate the different cell states within a cell type. C) PCA plots of the nuclei, which will give you the same plot every time you make this plot. However, since PCA is a linear method, you will typically need to make multiple PCA plots showing just two PCs at a time to appreciate the entire landscape of nuclei. All the plots originated from a dataset available through the SEA-AD consortium (Gabitto et al., 2023).*

**Clusters and cluster stability**

As stated previously, there's a biological distinction between cell states and cell types. Nonetheless, computational clustering of your cells can be beneficial for both tasks. A clustering method canonically partitions all your cells into one of many different clusters. These methods typically require you to input the number of clusters, either explicitly or implicitly. If you want to differentiate cell types, you can cluster your cells and look at which marker genes are highly expressed in each cluster. You often might purposely "over-cluster" your cells since you can combine clusters manually after the fact when investigating the marker genes. It's crucial to understand that the clustering procedure does not require marker genes, but you will need the marker genes to give the discovered clusters biological meaning. The typical clustering method in single-cell analyses are graph-based, such as Louvain or Leiden (Traag et al., 2019). These methods first compute a graph that represents a cell as a node, and two cells are connected if they have similar transcriptomic profiles.

Clustering serves a different purpose when analyzing cells of a single cell type, such as states of glial cells. These cells might have different transcriptomic profiles since they undergo state changes due to their environment or extracellular signaling. In these situations, clustering aims to find the subtle differences between cell states. Typically, differential expression or gene enrichment analysis is used to discover how the clusters differ. We will discuss this in detail later.

How do you pick the "correct" number of clusters? Some methods don't let you explicitly set the number of clusters, but instead have you set the "cluster resolution," which implicitly controls the number of clusters. This question is nuanced when working with cells of the same type but having different states. Many statistical tools have been developed to aid with these questions, but every tool is slightly imperfect in its own way. On the statistical end, methods related to data thinning are promising, where the dataset is split into two, the clustering is performed on one piece, and the quality of the clustering evaluated on the other piece of the dataset (Neufeld et al., 2024). However, such methods are difficult to deploy when many confounding variables, such as donor covariates, are involved. A more commonly used alternative is based on the stability of the clustering (Yu et al., 2022). These methods use the philosophy that a cluster's numeric stability over some user-generated randomness might hint at a biologically meaningful partitioning of the cells. For example, you can randomly select a group of genes and cluster the cells based on only those genes. After iterating this procedure multiple times where each time involves a slightly different set of genes, you can determine the best number of clusters as the one where the cell's cluster identity changed the least across the multiple iterations.

Lastly, we mention that there are also procedures that provide "soft" clusterings, where cells are instead treated as a weighted mixture of cell states. This is opposed to the abovementioned clusterings, typically called "hard" clusterings, where each cell is assigned to only one sub-state. Topic modeling is often used for this version of clustering (Carbonetto et al., 2022, 2023). While this modeling flexibility can be beneficial for finding differential pathways in your analysis, you will still have to deal with the question of picking the number of "pure" cell sub-states. While algorithms are available for the "hard" clustering methods discussed above, topic modeling does not benefit from methods that can assist in detecting the stability of "soft" clustering. As these methods develop further, we anticipate that additional algorithms will become available.

**Cell type annotation**

Once a dataset is visualized, the cell types and states can be annotated. Even if a dataset was enriched for a specific glial type, most sequencing experiments will result in at least a few cell types and several cell states. We demonstrated in the previous section that UMAP and t-SNE can both allow visualization of your cell types of interest because they often separate by gene expression. Historically, cells were annotated by visualizing cell type marker gene expression on a UMAP/t-SNE and then assigning cell type names to areas of the visualization where those markers were highly expressed. While the marker gene visualization is still useful, there are now computational tools available for cell type annotation.

Before discussing this, it is important to remember that the concept of "cell types" is quite nebulous. While some cell types are unequivocally different (there are many ways to distinguish a neuron from a microglia based on morphology, cellular function, transcriptomics, and spatial organization), some "cell types" are challenging to distinguish (Zeng, 2022). For example, the biological signature to define if a microglia is inflamed or in senescence is itself an active and evolving area of research (e.g. Ng et al., 2023; Saul et al., 2022; Vidal-Itriago et al., 2022). For our discussion here, we reserve the concept that two cells have different "cell types" if multiple biological modalities corroborate their differences. Here, "cell types" refer to the larger category of cells (e.g. neuron, astrocyte, oligodendrocyte, etc.). Otherwise, we reserve the word "cell state" to cells of the same type with slightly different cellular functions in the instantaneous moment defined primarily through transcriptomics. These "cell states" might often be described as subpopulations or phenotypes of a specific cell type. For example, we might call cells reprogramming differently due to stimuli or in different cell cycle stages as cells in different cell states (e.g. Batiuk et al., 2020; Chamling et al., 2021; Hammond et al., 2019; Matusova et al., 2023; Park et al., 2023).

The most common way to label cells by their cell type is through the help of marker genes. The marker genes for a particular cell type are selected to be uniquely (highly) expressed for only cells of this specific cell type relative to other cells. These genes are typically defined to have high sensitivity rather than high specificity. Usually, marker genes are defined by other labs, consortiums, or the literature. Once the marker genes of every cell type that may be in your experiment are organized, a computational procedure scores each cell for its enrichment for each set of marker genes to determine its likely cell type. We will discuss how a clustering of your cells can go hand-in-hand with this approach shortly.

This manual process of curating the marker genes of each cell type is fantastic for its transparency but can also be quite laborious. Hence, there are also computational methods to perform "label transfer" also known as using an existing annotation reference to label your cells. One advantage to label transfer is that these methods can facilitate the common labeling of both cell types and cell states in datasets (Aran et al., 2019; Xu et al., 2021). We note that these methods sometimes fall under the broader category of "data integration" methods and are sometimes repurposed to do batch correction. Our general recommendation is to apply label correction, visualize the data (as described before), and assess if the cell type labeling is satisfactory. See Abdelaal et al. (2019) for a more general discussion.

As a word of caution, first, as you use more granular cell states, the accuracy of these cell labels might not be as biologically meaningful. After all, other researchers defined these cell states using a possibly different experiment or biological model. Second, almost all the procedures described so far will struggle to identify cell types that exist in your dataset but not in the reference dataset. While most current label-transfer methods provide a "confidence score" that helps assess if there's a new cell type in your dataset, it is difficult to ascertain how reliable these computational procedures are. We strongly recommend that you spend time picking an appropriate reference dataset. A good reference dataset should be on a similar biological model, have a sequencing depth and sample size larger than your experiment (and so often come from consortiums), and have cell types labeled through procedures beyond single-cell RNA-seq data since transcriptomics only offers a partial view of a cellular phenotype. See Mölbert et al. (2023) for useful guidelines.

**Detecting differentially expressed genes (DEG)**

Differentially expressed genes (DEGs) allow the transition from computational analysis to new biological insight. DEGs are often calculated for two different purposes in an analysis. One is to determine what genes separate different cell types/states, and the second is to determine what genes have altered expression based on sample source or treatment. We will discuss both of these steps and how they differ in the next sections.

**Detecting DEGs for clusters:** This step enables identifying which genes separate two groups of cells. Typically, the groups are based on the clustering step discussed above. (In the next section, we will discuss how this differs from computing DEGs across different sources such as individuals or treatments.)

The most important guideline we offer is that DEGs should be identified using normalized but not batch-corrected/integrated gene counts. Using these further manipulated values artificially manipulates p-values and provides incorrect statistical inference. Some DEG algorithms (e.g., DESeq2, NEBULA) provide their own normalization, so starting from raw counts is most appropriate (He et al., 2021; Love et al., 2014). Other algorithms (e.g., MAST) expect log-transformed data, so normalized raw counts are appropriate (without further correction; Finak et al., 2015). If you use Seurat's SCTransform to normalize your dataset, the residuals or the log normalized corrected raw counts can be used for DEG analysis (Hafemeister & Satija, 2019). The most important thing here is not to use batch-corrected, PCA, denoised, imputed, or otherwise further transformed data in your DEG analysis. This message is illustrated in **Figure 4**.

Statistically, the batch-corrected or denoised gene expression matrix should not be used because most batch-corrected or denoised methods purposefully combine information across multiple genes together. This leads to interdependence of the gene expression because the genes are combined and no longer represented by their individual count values. This is the antithesis of an accurate DEG analysis since a truly differentially expressed gene should not contaminate the signal in a truly non-differentially expressed gene. This pitfall is well-documented and leads to the so-called "Type-1 error inflation" (Agarwal et al., 2020). While some methods have been developed to address such a scenario, we would advise you to use multiple DEG analyses and keep only the genes that multiple methods deem significant (Lin et al., 2024).

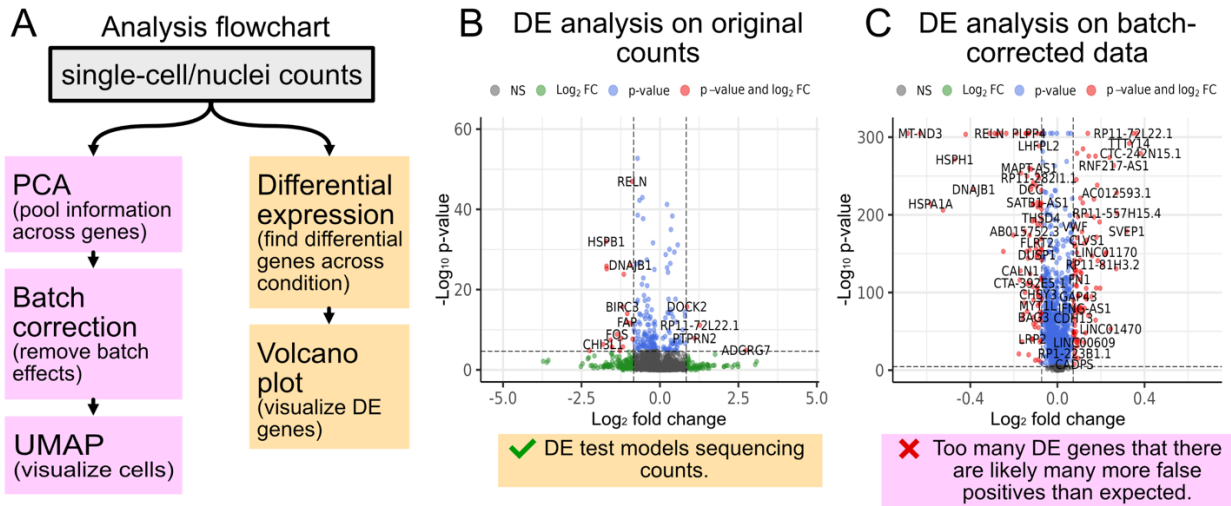

*Figure 4 caption: Workflow and example of DEG analysis. A) A flowchart illustrating that batch-correcting and visualizing the cells often uses a pipeline that is separate from computing the DEGs. B) A volcano plot when analyzing DEG genes, where the input data did not denoise genes by pooling information across the genes. We get a reasonable number of significant genes, marked by the horizontal dashed line denoting the multiple-testing threshold. C) A volcano plot when analyzing DEG genes, where the input data denoised the genes by pooling information across all the genes. We erroneously get too many significant genes. Notice that the y-axis scale is also dramatically higher. It is unlikely for most biological analyses that almost all the genes are contributing towards differences in the phenotype or treatment. The data here originates from the SEA-AD consortium when analyzing oligodendrocytes based on the pseudoprogression of donor's AD (Gabitto et al., 2023).*

Most of the DEG analyses we mentioned so far are only valid for the setting where the groups of cells are defined via experimental design. For example, a dataset might consist of DEGs between cells in differential experimental conditions or between cells that have undergone a treatment for differing lengths of time. This validity stems from the fact that the groups of cells were defined independently from the gene expression itself. The statistical validity of the p-value gets less reliable when you begin to deviate from this rule. The most common deviation occurs when you use the single-cell dataset to cluster the cells into different sub-states and then use the same exact single-cell dataset again to test for significant DEGs (Lähnemann et al., 2020). Unsurprisingly, this produces extremely high levels of DEGs since the clustering method partitioned the cells into different sub-states based on the differences in gene expression (Zhang et al., 2019). This "double-dipping" has been the focal point of many new statistical methods (Vandenbon & Diez, 2020; Zhang et al., 2019). While it is possible to use clusters defined by gene expression to then find DEGs with valid p-values using sophisticated statistical methods, it is generally recommended to use a reference dataset to cluster so that DEGs can be identified in the dataset in a more independent way (Lähnemann et al., 2020). As mentioned above in the section on cell type labels, it is quite important to identify a well-matched reference dataset from which to define clusters, and one may not be available in all instances. We anticipate that as the number of glial single-cell datasets grows appropriate reference datasets for clustering will become more available.

**DEGs for group differences**: If the scientific question involves groups of donors (i.e. case vs. control, treated vs. untreated, young vs. aged), then the previous DEG methods do not offer the most insightful results. This is true generally, but especially for the human population. This is because many DEG methods were designed for cell lines and clonal mice but not for human donors, where there are considerable donor covariates, such as age and sex. For instance, if you were studying glial cells in Alzheimer's Disease (AD), you might compare the donors with a high AD pathology burden to those with a low AD pathology burden based on the single-nucleus RNA-seq data from all the donors in the cohort.

It is important to understand why the typical DEG analysis we mentioned previously might not be appropriate for cohort-level analyses. First, there is considerable human variation that a cohort-level analysis needs to account for. For example, when analyzing data from human cohorts where some donors were diagnosed clinically with AD, it's common to find that almost every gene separates the AD donors from control donors. However, some of these differences might be driven by age or sex or other confounds that are not the biological aspect that you are studying. Second, when studying the human population, the number of cells and the number of donors play a different role. Consider two hypothetical datasets, each with one million cells. One dataset sequenced two donors with half a million cells each. The other dataset sequenced a thousand donors with a thousand cells each. While the former dataset offers a more thorough picture of all the cells in two donors, the latter offers findings that generalize better to the human population. Third, the previous DEG methods would categorize the cells into two groups based on the phenotype of the originating donor. These methods would not be able to account for within-donor variation. This oversight could lead to inaccurate results.

How should we analyze cohort-level data then? The most used method is to "pseudobulk" the single-cell RNA-seq data based on the donors. This means you compute the summed count expression matrix among all the cells of a particular cell type for each donor, and then you perform a DEG analysis originally designed for bulk RNA-seq data. Typically, this is done using bioinformatics packages such as DESeq2 or Limma (Love et al., 2014; Ritchie et al., 2015), which can adjust for donor covariates. This procedure is often called "pseudobulking" because we are computationally emulating sequencing bulk data from each donor. Pseudobulking may be counter-intuitive since you sequenced single-cell data only to analyze bulk-level data. However, the advantage of this procedure is that you first label your single-cell dataset based on their cell types and then perform pseudobulk analysis on cells of one cell type or cell state specifically. Using your clusters provides the power of single-cell datasets to identify subpopulations but pseudobulk statistically uses donor covariates appropriately.

While pseudobulking is considered a "gold standard" in a cohort-level analysis, it potentially has low power due to its failure to account for within-donor gene expression variation. Furthermore, since tools used to analyze psuedobulk were designed for bulk RNA-seq studies where RNA input is matched, there is not always a built in control for cell number. It can also yield false positives in certain cases where the within-donor variation is large, relative to the between-donor variation. Specific methods have been developed to analyze cohort-level data at the single-cell level (e.g. He et al., 2021; Lin et al., 2024). The same warning that we mentioned previously about pooling information across genes for DEGs still holds. Mixed-effect models adjust for donor covariates when analyzing one gene at a time. The sparsity in single-cell RNA-seq data creates challenges in this adjustment when complex relationships exist between the donor covariates and the gene expression. Embedding methods like eSVD-DE are better able to provide reliable p-values while accounting for donor-level covariates.

Overall, note that increasing the number of cells for one donor allows profiling all the different cell states present in that donor with high fidelity, and increasing the number of donors allows the dataset to generalize more readily to the human population. When planning your experimental design to study a cohort of donors, it's important to strike a good balance between the number of cells per donor, the sequencing depth for each cell, and the number of donors. Suppose an existing dataset of a similar cohort already exists. In that case, a power calculation can be made to determine how many donors you might need to recruit into your cohort (reviewed in Jeon et al., 2023). Be aware that your own study might have a different number of cells per donor or sequencing depth. There has yet to be a consensus on how to perform cohort-level single-cell analyses, so it may be useful to discuss an analysis plan with a fellow statistician.

**Biological function correlates**

One of the most applied analysis tools after computing the DEGs is pathway analysis. This is typically done as gene set analysis or gene set enrichment analysis. There are significant differences between these two approaches. Gene set analysis (GSA) is applied when a user takes a list of significantly different genes and uses a tool like DAVID or Panther to perform a Fisher's exact test to identify putative biological pathways associated with that particular list of genes (Mi et al., 2019; Sherman et al., 2022). You can interpret this analysis as asking which biological pathways are most likely associated with the genes identified. Gene set enrichment analysis (GSEA), in contrast, takes a ranked list by fold change of all genes present in the analysis, and utilizes the directionality of their change in addition to the amount of change to assess pathways where significant numbers of genes are changing in the same direction. This analysis does not require the genes themselves to be significantly altered, but if a large number are shifting together in the same direction then a pathway will be called significant. You can interpret the results of this analysis as providing biological pathways that are changing in either a positive or negative direction based on gene expression shifts as a result of the condition of interest. Since each of these pathways analysis approaches utilizes gene expression in a different way, the results are not perfectly aligned. GSEA provides the opportunity to identify pathways that are shifting because of coordinated gene expression changes even if the gene expression changes are not themselves significant, whereas GSA only utilizes significantly expressed genes and does not take directionality or magnitude into account.

GSA is a useful tool; however, interpretation of these pathways and the links to genes can be less clear than GSEA, particularly if the user does not separate their list of genes into those positively and negatively regulated by the condition of interest. Thus, when applying GSA it is important to remember that only the significantly different genes are taken into account, and directionality is only taken into account so far as the user splits their lists apart. Magnitude of change is also not assessed in this analysis. GSA is best used when the list of genes is long. If a list of differential genes is on the shorter side, for example having only 20 genes, the reliability is likely low.

GSEA provides additional capability beyond GSA because it utilizes both magnitude and directionality of change to identify pathways that are altered in the dataset. Although GSA is simpler because multiple websites exist where lists of gene names can be supplied and pathways retrieved, we recommend the use of GSEA because of the additional information that is utilized in identifying the pathways.

When identifying biological pathway correlates, it is also good to consider that each of the commonly utilized databases (e.g. GO, KEGG, Reactome, Wikipathways) is based on a specific set of scientific papers curated by one or a group of individuals (Agrawal et al., 2024; Ashburner et al., 2000; Kanehisa, 1997; Milacic et al., 2024; The Gene Ontology Consortium et al., 2023). These databases are biased by what is known about biological pathways in certain fields with which the curators are familiar and may not be accurate for all cell types or situations. One of the ways that scientists have attempted to overcome this bias is to utilize multiple databases and then identify the common pathways across databases rather than selecting a single database from which to draw results. Ideally, this makes the resulting pathways identified more reliable as they are shared beyond the bias of a single database. While the databases themselves will continue to improve, they will always be inherently biased by what is already known in science. By combining the results of databases and reporting the common themes, confidence is higher that the pathways identified represent the biology seen in these gene expression shifts.

**Inferring gene regulatory networks or ligand-receptor interactions and WGCNA**

After one defines the differentially expressed genes in the dataset, there are many optionally applicable downstream analysis options depending on the scientific questions to be answered. This could include inferring gene regulatory networks (Aibar et al., 2017; Badia-i-Mompel et al., 2023; Yuan & Duren, 2024). Tools are also available to detect sets of genes that change together (similar to the GSEA concept described above) but then to correlate that coordinated shift in expression with known transcription factors that drive gene expression. This allows the identification of gene regulatory networks. One tool commonly utilized in bulk RNA-seq publications is weighted gene network correlation analysis (WGCNA; Langfelder & Horvath, 2008, 2012). Similar to GSEA, this method detects genes in the dataset that move up or down in a correlated or coordinated manner. There are methods that allow WGCNA to be applied to single-cell RNA-seq datasets (Morabito et al., 2023), and other tools that expand the capability (Lu & Keleş, 2023; Su et al., 2023). Other tools utilize databases that identify ligand-receptor interactions in different fields of biology to infer the way cells may be communicating with each other based on the gene expression in the dataset (Armingol et al., 2021; Jin et al., 2021; Wilk et al., 2024; Xie et al., 2023). All of these methods utilize the DEGs detected in the dataset and rely on databases that already exist so the choice of database and the way DEGs are calculated are important inputs.

**Trajectory analysis**

Typically, when isolating the glial cells in your single-cell datasets, one might care about how cells change over time. Moreover, one might hypothesize that different glial cells undergo different transcriptomic changes. For example, these changes might be due to experimental perturbation in the wet lab or cellular response due to increasing neurodegenerative burden. Whichever the case might be, a trajectory analysis might provide insight into what exactly is changing with respect to time. What makes this complicated in glial analyses, however, is that these methods are often designed for developmental systems where there's a clear development of vastly different cell types. Hence, your glial analysis is less likely about "development" but more likely about "responding" (i.e., how the glial cells gradually shift). This distinction affects how likely it is that the trajectory analyses will provide meaningful insight.

Trajectory analyses, broadly speaking, fall into two categories. The first category requires you to specify the "start" and/or "end" of the trajectory. If you suspect multiple trajectory routes, some methods require you to enumerate each route's start and end. Then, these methods find paths through the transcriptome space (often the PCA embedding of the normalized gene expression data) that go through the high-density regions between each start and end (Street et al., 2018; Trapnell et al., 2014). The second category instead uses a molecular model to dictate how cells should change over time and typically requires additional annotations of your gene expression matrix (La Manno et al., 2018). Neither category is strictly preferable. While the first category requires you to label the routes roughly, the second requires you to trust that the molecular model is appropriate for your biological system.

A few downstream analyses are commonplace once you've performed a trajectory analysis. One is to compute the pseudotime of each cell in the dataset. We use the word "pseudotime" since most trajectory methods do not explicitly account for how time affects gene expression. That is, there are no meaningful units to pseudotime. Biologically, we typically interpret the pseudotime as a mathematical proxy of how far along a cell is in responding to a perturbation or a disease's burden. The second downstream analysis is to find the order of cascading genes. These are genes that are highly expressed at different times along any trajectory. The third downstream analysis is to identify the branch points. If a glial dataset has multiple trajectory routes, one might be interested in which cell sub-state was the "last" sub-state before the cells had to "make a decision" on which later state would be its endpoint. Note that all three downstream analyses are often specific to the trajectory analysis. The exact calculation for each downstream analysis will likely change depending on the trajectory inference applied.

**Figure 5** demonstrates one common weakness of existing trajectory methods when analyzing subtle glial shifts. While many trajectory methods were designed for studying early embryonic development where cells are changing their cell types, for example a stem cell developing into a neuron, you might be instead studying how microglia change their cell states. This means there is dramatically less distinction between the cell states than most trajectory methods require. In Figure 5, we demonstrate this, where our suggestion is to downsample the number of cells, apply the trajectory method on the dataset with less cells, and repeat this procedure multiple times to build a consensus among the random subsetting of the dataset.

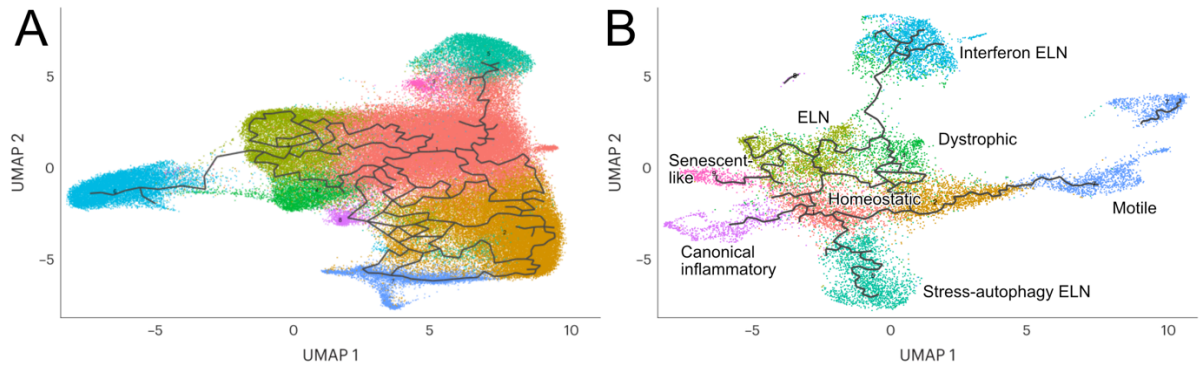

***Figure 5 caption***: *Trajectory analyses. A) Analyzing the trajectories of microglia, where the lack of nuclei separation among the clusters yields overly complex and likely irreproducible trajectories. B) Analyzing the trajectories of the same dataset, but now downsampling the nuclei. By having less nuclei, the trajectories are less complex, and this downsampling procedure is redone many times (not shown) to ensure that the trajectories do not vary with the downsampling randomness. Both figures are analyzing the same data of microglia from Prater et al., 2023 using Monocle. The nuclei are colored by their cell state cluster.*

A final word of caution is that the vast majority of trajectory analyses are exploratory in nature. These analyses often do not claim any causality of how cells are responding. In fact, many authors argue that the regulatory network is a more meaningful biological representation of a cell's identity (Kamimoto et al., 2023). Validation of trajectory analysis findings is fruitful. The most common strategy is to find the cascading genes along a trajectory and perform a GSA on those genes. Another is to correlate the pseudotime with a biologically and independently derived score of the cells. For example, if you are analyzing cells from a human cohort studying a neurodegenerative disease, you might separately construct a score for the neurodegenerative burden in each glial cell. Then, you would assess how correlated pseudotime is with this neurodegenerative burden score.

**Composition analysis**

The last topic we will cover is compositional analysis. After all, the benefit of collecting single-cell data is that we can annotate individual cells by their cell type. We can then compute the proportion of cells across different treatment conditions in wet bench experiments or across different donors with varying phenotypes. This allows us to ask questions about the changing composition of cells. For example, we might hypothesize that certain microglia states become more prevalent as the AD burden increases. This hypothesis is fundamentally a compositional question, as we are less interested in which pathways or genes are being activated. Hence, the previous DEG methods are not applicable.

While you might be tempted to compare the measured proportion of cell types across the different conditions to answer this question, we warn you of a few caveats. First, cell proportions sum up to 1, so an increase of one cell composition necessarily means the other cell type proportions decrease if all other biological factors were held fixed. This demonstrates the complex nature of a compositional analysis. You might wonder if activated astrocytes became more prevalent, or if other astrocyte states died. While both scenarios might mathematically yield the same compositional proportions, they have foundationally different scientific meanings. Second, the cell types or states are estimated, so you have a variable estimate of the cell type proportion. This might result in your analysis being overly confident about a change in proportion when, in reality, the estimate of proportion itself is noisy.

Specific tools (scCODA and Cacoa) can help perform this compositional analysis (Büttner et al., 2021; Petukhov et al., 2022). In short, these methods often elect one cell type or cell state to the "reference class," and assume that two treatment conditions do not affect the cells in this reference class. Investigators need to rely on the activated genes of each cell type or cell state to biologically make a meaningful choice on which one should be the reference class. Additionally, be aware that the meaning of "samples" differs when doing a compositional analysis. Whereas in the DEG or trajectory analysis, the more cells sequenced yield more power, a compositional analysis's power is limited by how many donors or replicates there are. For example, a perturbation experiment performed across two different conditions on a cell line with only two replicates for each condition will only have four samples (two in each of the two conditions). This design means there might not be enough statistical power to detect significant changes in cell composition.

**Conclusion**

Single-cell technologies and the analyses that can be applied to them have rapidly evolved and now are widely available. Here we call out common pitfalls in terms of batch analysis, dataset visualization, and DEG detection. We also provide an outline of experimental design and analysis considerations with a focus on how these decisions may differ when working with glia.

Like all science, quality single-cell datasets begin with careful experimental design. The 'omics technology needed to answer the scientific question as well as a careful consideration of batch effects that may be introduced in the data are key. Glial biologists may also consider whether to enrich for their cell type of interest to provide better power in detecting cell states within the population.

Once a dataset is collected, analysis begins with quality control. Quality control consists of thresholding to remove empty droplets and the choice of applying ambient RNA correction and doublet detection. Doublets may be removed, or simply labeled in a dataset, where the label allows for later removal or ignoring of the effects of those droplets in downstream analysis.

Following quality control, the downstream analyses take two routes. The first algorithms to be applied are in the lefthand lower portion of Figure 1 in Dataset visualization. Normalization is an important step in detecting the biological signal of gene expression and removing technical noise. One then needs to determine whether batch correction is needed for the dataset, and if so, which batch correction method performs the best. Often, this requires running more than one batch correction method and comparing the resulting dataset visualization. As we note in the introduction, single-cell datasets often require visualization for informed decision-making which is a unique difference compared to standard statistical approaches to data. After batch correction, the dataset can proceed down PCA dimension reduction to visualization using UMAP or t-SNE to clustering, cluster annotation, composition analysis, and trajectory analysis among other downstream analyses detailed above. These analyses typically result in a visualization of the dataset using either UMAP or t-SNE.

The second analysis route after quality control results in differential gene expression and analyses downstream of that detection. These analyses are not appropriate to perform on batch corrected data, so are considered a separate analysis stream. Some DEG detection algorithms require normalized data to operate effectively, but the first thing to determine is what form of data (normalized or raw counts) the algorithm expects. From there, scientists can detect cluster-specific DEGs that utilizes the clustering done in analysis stream 1 but doesn't rely on the batch corrected dataset itself. When group-level questions are involved (how do two treatments differ, or do cases differ from controls) then pseudobulk is the gold standard for DEG approaches. There are newer algorithms that allow for modeling the complexity of single sample sources for all the single cells that are also effective for answering group level questions. Once DEGs are defined, then biological pathway enrichment, cell-cell communication, gene regulatory networks, and other downstream analyses can be added.

Together, we hope that the information provided will allow the advance of glial biology in a rigorous way.